\documentclass[aps,prl,twocolumn,showpacs,floatfix,superscriptaddress]{revtex4-1}
\usepackage{graphicx}
\usepackage{color}
\usepackage{amsmath}
\usepackage{amssymb}
\usepackage{amsfonts}
\usepackage{comment}
\usepackage{bm}
\usepackage{braket}
\definecolor{scarred}{rgb}{0.75,0.0,0.0}
\usepackage{hyperref}
\hypersetup{
colorlinks=true,final=true,
        linkcolor=blue,
        citecolor=blue,
        filecolor=blue,
        urlcolor=blue,
}
\newcommand{\jl}[1]{\textcolor{blue}{#1}}

\begin{document}

\title{A photo-induced strange metal with electron and hole quasi-particles} 
\author{Nagamalleswararao Dasari}\email{nagamalleswararao.d@gmail.com}
\affiliation{Department of Physics, University of Erlangen-Nuremberg, 91058 Erlangen, Germany}
\author{Jiajun Li}
\affiliation{Department of Physics, University of Erlangen-Nuremberg, 91058 Erlangen, Germany}
\author{Philipp Werner}
\affiliation{Department of Physics, University of Fribourg, 1700 Fribourg, Switzerland}
\author{Martin Eckstein}\email{martin.eckstein@fau.de}
\affiliation{Department of Physics, University of Erlangen-Nuremberg, 91058 Erlangen, Germany}

\begin{abstract}
Photo-doping of Mott insulators or correlated metals can create an unusual metallic state which simultaneously hosts hole-like and electron-like particles. We study the dynamics of this state up to long times, as it  passes its kinetic energy to the environment. When the system cools down, it  crosses over from a bad metal into a resilient quasiparticle regime, in which quasiparticle bands are formed with separate Fermi levels for electrons and holes, but quasiparticles do not yet satisfy the Fermi liquid paradigm. Subsequently, the transfer of energy to the  environment slows down significantly, and the system  does not reach the Fermi liquid state even on the timescale of picoseconds. The transient photo-doped strange metal  exhibits unusual properties of relevance for ultrafast charge and heat transport: In particular, there can be an asymmetry in the properties of electrons and holes, and strong correlations between electrons and holes, as seen in the spectral properties.
\end{abstract}

\maketitle

{\textit{Introduction --}} Ultrafast laser excitation provides a new avenue for manipulating correlated quantum states in condensed matter \cite{Claudio2016,Basov2017}, and has led to intriguing observations such as light-induced superconductivity and metastable hidden phases \cite{Fausti2011,Kaiser2014, stojchevska2014,mitrano2016}. An interesting direction in this context is the optical manipulation of Mott insulators (MIs), which are the parent compound for a variety of  complex states  \cite{Lee}. Because chemical doping leads to correlated metallic, pseudo-gap and super-conducting phases, photo-doping, i.e., an optical excitation which simultaneously generates electron-like and hole-like carriers, has early on been identified as a potential route for materials control \cite{Iwai2003, Wall2011, Okamoto2007, Okamoto2010, Okamoto2011, Mitrano2014, Miyamoto2018, Petersen2017, Sahota2019}. 
A short pulse can almost instantly transform the MI into a hot metallic state. The presence of carriers is demonstrated by a Drude peak in the conductivity, and because a large Mott gap prevents a rapid carrier recombination through electron-spin, electron-phonon, or electron-electron scattering \cite{Strohmaier2010, Sensarma2010, Lenarciz2013, Lenarciz2014, Lenarciz2015, Eckstein2011}, these carriers can last up to picoseconds. On this timescale, the energy transfer to the environment may then establish a {\em cold photo-doped state} before recombination of the photocarriers. More generally, applying the same protocol to an already doped MI will generate a strange metal with unequal densities of electron-like and hole-like carriers.
The properties of such a correlated liquid, and in particular how they differ from the chemically doped state, remains an intriguing question.

For very large photo-doping, e.g., $\eta$-pairing superconductivity has been predicted \cite{Rosch2008, Peronaci2020, Li2019}, but already the normal state properties pose fundamental questions, as doped MIs are often strange metals with non-Fermi liquid properties \cite{Deng2013, Georges2013, wang2019, Legros2019, Daou2009, Werner2008, Parcollet1999,Vucovic2015, Pakhira2015}. For example, quasi-particle excitations can be observed at temperatures well above the range of validity of Fermi-liquid theory, $T_{\rm FL}$, before they disappear in the bad metallic regime around $T_{\rm MIR}$ \cite{Deng2013}. Such resilient quasi-particle excitations dominate the transport properties of doped MIs in the intermediate temperature regime $T_{\rm FL} \lesssim T \lesssim T_{\rm MIR}$. Out of equilibrium, additional fundamental questions arise: Can there be a Fermi liquid at all with both electron and hole quasiparticles at {\em different} Fermi energies? In principle, such a two-species Fermi liquid is possible if the number of each species is separately conserved. In the present case, however, this conservation is only valid up to a given timescale, while the time for the formation of a Fermi liquid is at least  beyond the range of previous numerical simulations \cite{Martin2013,Sayyad2016}. Another interesting question concerns the particle-hole asymmetry: If the initial state is already doped from the outset,  photoexcitation will lead to a photo-doped state with an imbalance in the electron and hole doping. As properties of carriers in a doped Mott insulator strongly depend on the doping level, these two types of carriers could exhibit different properties. If so,  this should not only influence the charge transport properties, but also the rather little explored question of thermo-electric properties on the ultrafast timescale. 

While dynamical mean-field theory (DMFT) \cite{Georges1996} is in principle ideally suited to study the Mott phase, previous simulations simply could not reach sufficiently long times to systematically examine the cold photo-doped states \cite{Martin2013, Sayyad2016, Peronaci2020}. Here we use a systematic and convergent truncation of the Kadanoff-Baym equations within DMFT to reach about $50$ times longer simulation times \cite{Schueler2020,Stahl-TBP}. This allows us to study the properties of the photo-doped MI up to thousands of hopping times, corresponding to a picosecond timescale if the bandwidth is in the eV range.

{\textit{Model and Method  --}}  We simulate the relaxation dynamics of doped Mott insulators by considering the single-band Hubbard model,
\begin{equation}
{\cal{H}} = -J\sum_{\langle ij \rangle,\sigma} c^{\dagger}_{i\sigma} c^{\phantom{\dagger}}_{j\sigma} 
+ U \sum_i n_{i\uparrow} n_{i\downarrow} - 
\mu \sum_{i\sigma} n_{i\sigma}.
\label{eq:latt}
\end{equation}
The first term in the above Hamiltonian represents the electron hopping between nearest-neighbor lattice sites, and the second term is the onsite Coulomb interaction between electrons of opposite spin. The chemical potential $\mu$ fixes the total number of particles. In equilibrium, the system is a Mott insulator at half filling $n=\langle n_{i\uparrow} + n_{i\downarrow}\rangle=1$
and $U \gg J$, and turns into a strongly correlated metallic state upon doping. 
Without loss of generality, we focus on states which are initially half-filled or electron-doped, where $\delta=n-1$ quantifies the initial chemical doping. We solve the Hubbard model on a Bethe lattice using non-equilibrium dynamical mean-field theory \cite{Aoki2014}. The interacting local Green's function $G(t,t')$ is obtained by solving the auxiliary quantum impurity problem on  the L-shaped Kadanoff-Baym contour using the non-crossing approximation \cite{Martin2010}. In our calculations, we fix the hopping energy scale $J=1$, so that bare bandwidth is $W$=4$J$. Furthermore, we set $\hbar$ = 1 and measure time in units of $\hbar/J$. Unless otherwise stated, we fix the Hubbard interaction to $U=8$. In addition to the Hamiltonian \eqref{eq:latt}, a thermal bath is coupled to each site to simulate the energy dissipation from the electrons to other degrees of freedom \cite{Martin2013}. The bath is incorporated diagrammatically with a self-energy $\Delta_\text{bath}(t,t')$ = $\lambda G(t,t') D_\text{bath}(t,t')$ which corresponds to a Holstein type coupling of strength $\lambda$ to a bath of bosonic degrees of freedom at given temperature $T_f$ and  a linear (Ohmic) density of states, $D_\text{bath}(\omega)$=$\frac{\omega}{\omega_c} e^{-\omega/\omega_c}$ up to a cutoff energy $\omega_c=0.2$. $D_\text{bath}(t,t')$ is the corresponding bath propagator. The DMFT equations are analogous to Ref.~\cite{Martin2013}, but they are solved using a convergent truncation of the memory integrals in the Kadanoff-Baym equations. Details of the implementation are given in the appendix. 

\begin{figure}[tbp]
\includegraphics[scale=1.3]{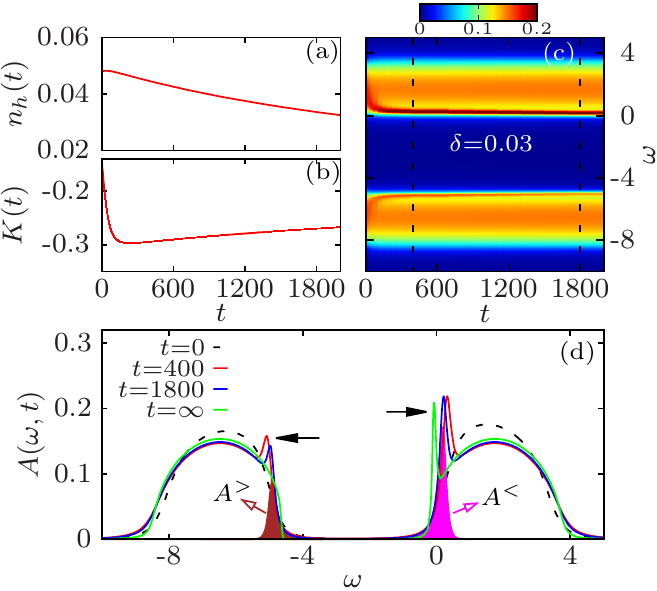}
\caption{Dynamics of the photo-doped state at $\delta=0.03$. (a) Photo-doping level $n_h(t)$. (b) Kinetic energy $K(t)$. (c) Intensity map of the  single-particle spectral function $A(\omega,t)$. (d) Line cuts of panel a), showing the spectral function at selected times. Black arrows indicate the electron and hole quasi-particle states in the spectral function. The shaded area labelled with $A^>$ and $A^<$ 
shows the unoccupied (occupied) density of states in the lower (upper) Hubbard band at time $t=1800$.}
\label{fig:fig1}
\end{figure}

{\textit{Setting --}} We initially prepare an electron-doped system of filling $n=1+\delta$ at a very high temperature $T_i=2$ (in the bad metal regime), and couple it to a heat bath at lower temperature $T_f=0.05$ for times $t>0$. This protocol serves to generate simultaneous hole and electron doping  (see below) through thermal excitation $T_i$  rather than through photo-excitation. We nevertheless refer to the transient state as {\em photo-doped}, because we focus on its long-time behavior, while details of different carrier generation protocols quickly become irrelevant \cite{Martin2013,Li2020}. The system can be characterized by  the time-dependent single-particle spectral function $A(t,\omega)=-\frac{1}{\pi} \text{Im}\int^{t_{max}}_0 ds \,G^{R}(t,t-s) e^{-i\omega s}$ and the corresponding occupied density of states $A^<(t,\omega)$ and unoccupied density of states $A^>(t,\omega)=A(t,\omega)-A^<(t,\omega)$, see Fig.~\ref{fig:fig1}d. The spectra feature well-separated upper and lower bands, so that a time-dependent electron and hole doping value can be obtained from the integrated occupied density of states in the upper band, $n_e(t)=\int_\text{UHB} d\omega A^<(\omega,t)$, and the integrated unoccupied density of states in the lower band, $n_h(t)=\int_\text{LHB} d\omega A^>(\omega,t)$, respectively. In a low-temperature equilibrium state, $n_h(t)=0$ and $n_e(t)=\delta$, but in the transient state the system has both hole doping $n_h>0$ and electron doping $n_e>\delta$, with $n_h=n_e-\delta$ of the order of a few percent.

{\textit{Results --}} A first picture of 
the formation of a cold photo-doped state is given by the behavior of the photo-doping level and the kinetic energy. The decay of the photo-doping $n_h(t)=n_e(t)-\delta>0$ due to carrier recombination is slow for all times, as expected for $U\gg J$ and a large Mott gap (Fig.~\ref{fig:fig1}a). The kinetic energy instead drops quickly at early times, as energy is transferred from the electrons to the bath and the cold photo-doped state is formed, and then follows the slow dynamics of the carrier recombination (Fig.~\ref{fig:fig1}b). The spectral function (Fig.~\ref{fig:fig1}c and d) shows just two bands of incoherent excitations above and below the Mott gap immediately after the temperature quench.  Within a few hundred hopping times, a narrow band emerges at the upper edge of the Mott gap, which is reminiscent of the quasi-particle band of the electron-doped Mott insulator in equilibrium. At the same time, a similar (smaller) hole quasi-particle band emerges at the lower edge of the Mott gap.  We emphasize that these quasiparticle bands emerge as a consequence of (photo)-doping, and are therefore different from the fine-structure at the edge of the Hubbard bands which can be seen in equilibrium at half filling closer to the metal-insulator transition \cite{Lee2017}. At late times, these quasiparticle peaks slowly evolve, which is a consequence of the reduction of the effective temperature (see discussion below), and the decay of the photo-doping.  The behavior is qualitatively the same for different values of the electron-phonon coupling $\lambda$ (the results are for $\lambda=0.5$ unless otherwise stated). Previous DMFT simulations of photo-doped Mott insulators \cite{Martin2013} could only observe the onset of the formation of these quasiparticle bands, and have mostly focused on the half-filled case. With the long-time simulations, we can now address in more detail the relaxation behavior and the spectral properties of the state. We first analyze these properties at the latest time ($t=2000$), before discussing the relaxation dynamics.

\begin{figure}[tbp]
\includegraphics[scale=1.3]{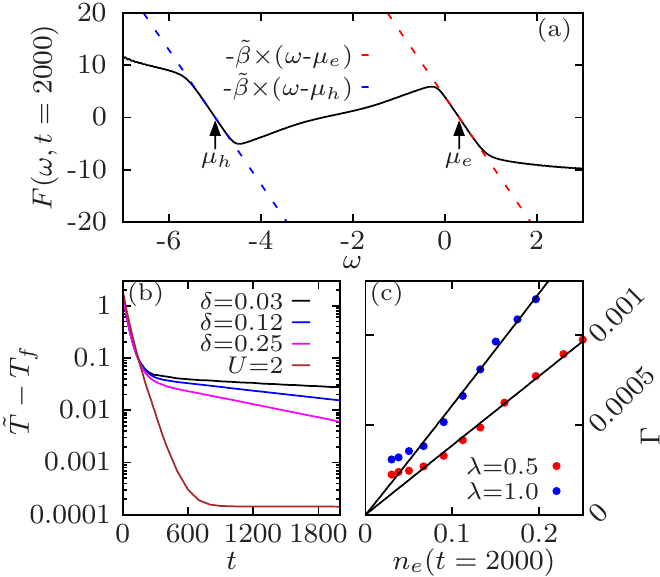}
\caption{Distribution function of the transient state for $\delta$=0.03. (a) $F(\omega,t)$ at $t=2000$. Dashed lines show linear fits around the location of the electron and hole quasi-particle peaks. Vertical arrows indicate the electron and hole Fermi-levels. (b) Difference of the effective temperature $\tilde T=1/\tilde\beta$, obtained from fits like in panel a) at different times,  and the bath temperature $T_f=0.05$, as a function of $t$ for different values of $\delta$. (c) Exponential decay rate of $\tilde T-T_f$ obtained from the curves in b) at $t=2000$, plotted against the photo-doping  $n_{e}(t)$ at the same time. }
\label{fig:fig3}
\end{figure}

The nature of the transient state is further explored by looking at the distribution function $f(\omega,t)=A^{<}(\omega,t)/A(\omega,t)$.  In an  equilibrium state, $f(\omega)$ is just the Fermi function. It is therefore convenient to analyze the function $F(\omega,t)$=$\log \left[A^{<}(\omega,t)/A^{>}(\omega,t) \right]$, which becomes $F(\omega)=-\tilde \beta(\omega-\tilde \mu)$ if the system is in a quasi-thermal state in which the distribution is given by a Fermi function with inverse temperature $\tilde \beta$ and Fermi level $\tilde \mu$. In the transient state, $F(\omega,t)$ has such a linear behavior in the frequency range of the electron and hole quasi-particle peaks (Fig.~\ref{fig:fig3}a). In the remaining frequency range, either the spectral weight is small (gap), or the occupation is close to zero or one. The distribution function therefore shows that the photo-doped Mott insulator at long times is well described by two separately thermalized subsystems of electron-like and hole-like quasiparticles. From the linear fit of $F(\omega,t)$ to $-\tilde \beta_{e,h}(\omega-\mu_{e,h})$, we can extract the inverse effective temperature $\tilde{\beta}$, which turns out to be approximately the same for electrons and holes, and the separate Fermi levels  $\mu_{e}$ and $\mu_{h}$ of these degrees of freedom (arrows in Fig.~\ref{fig:fig3}a).

\begin{figure}[tbp]
\includegraphics[scale=1.3]{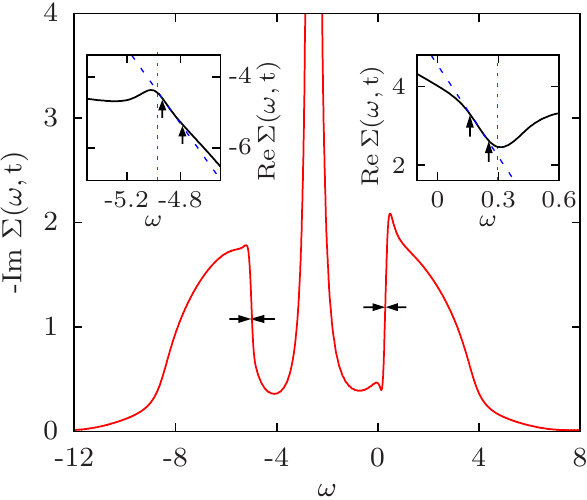}
\caption{Imaginary part of the self-energy in the transient state for $\delta$=0.03 and $t$=2000. The horizontal arrows indicate the electron and hole Fermi levels. Insets: Real part of the self-energy near the electron and hole Fermi levels (red dashed lines). Blue dashed lines show the linear fit to the real-part of the self-energy. The vertical arrows indicate the energies where the deviations from the linear behavior occur.}
\label{fig:fig2}
\end{figure}

At zero temperature the Fermi surface separates occupied energy states from un-occupied states with a step-function \cite{Dugdale_2016}. In the present case, this step is broadened, and whether the electron and hole liquids can be considered as a Fermi liquid must be answered by the spectral properties, in particular, the self-energy. In a Fermi liquid, the scattering rate of the dressed quasi-particles near the Fermi-surface is determined by the single-particle self-energy $\gamma(\omega)=-\operatorname{Im}\hspace{0.03cm}\Sigma(\omega) \sim C\omega^2+C'T^2$ \cite{Landau}. We therefore analyze the self-energy around the positions $\tilde\mu_{h,e}$  of the electron and hole quasi-particle peaks in the spectral function (Fig.~\ref{fig:fig2}); $ \Sigma(t,\omega)$ is obtained from the Dyson equation, $ \Sigma(t,\omega)=G_0(t,\omega)^{-1}-G(t,\omega)^{-1}$. At the Fermi-levels, $\gamma(\omega,t)$  has a linear behavior rather than a quadratic form, which indicates that the transient state does not belong to the Fermi-liquid regime, although quasiparticle peaks are well defined in the spectral function. This is nevertheless analogous to the resilient quasi-particle regime \cite{Deng2013} in equilibrium. The real part $\operatorname{Re}\hspace{0.03cm}\Sigma(\omega,t)$ of the self-energy shows a linear behavior around the quasi-particle peaks, which, however, does not extend up to the Fermi levels. Deviations from the linear behavior are observed at two distinct energy scales, see vertical arrows in Fig.~\ref{fig:fig2}. The change in the slope of the linear behavior should lead to the appearance of characteristic `kinks' in the quasi-particle dispersion \cite{byczuk2007}.

\begin{figure}[tbp]
\includegraphics[scale=1.3]{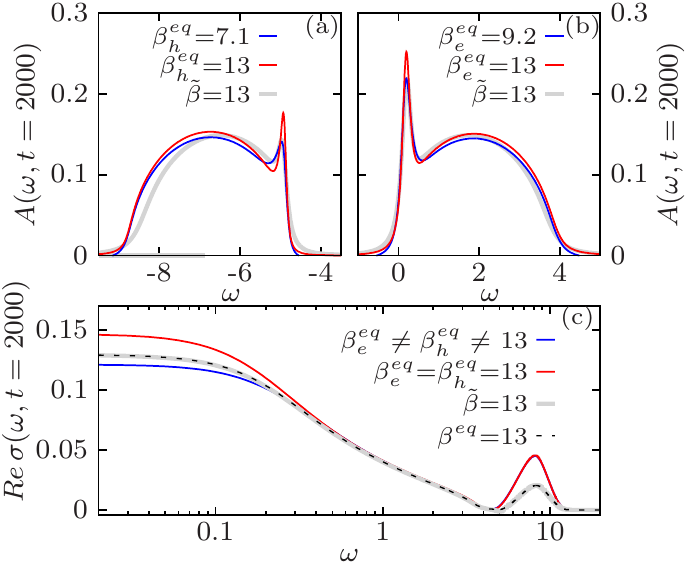}
\caption{Comparison between the photo-doped state at $\delta=0.03$ and $t=2000$, where $n_e=0.062$ and $n_h=0.032$, and chemically doped states: (a) Lower Hubbard band of the photo-doped state (light grey) and for an equilibrium  system with filling $n=1-n_h$, at $\beta=13$  (same temperature $1/\tilde \beta$ as the photo-doped state) and $\beta=7.1$. (b) Comparison of the upper Hubbard band to electron-doped systems. (c) Optical conductivity $\sigma(\omega)$ of the transient state (bold light grey), of an equilibrium system with filling $n=1+n_h+n_e$ and $\beta=13$ (black dashed line), and sum of the conductivities of systems with $n=1-n_h$ and $n=1+n_e$ at $\beta=13$ (bold blue and red line). }
\label{fig:fig4}
\end{figure}

To further understand the nature of the photodoped state and, in particular, the origin of the resilient quasi-particles, we next compare the transient state with chemically doped states. In Fig.~\ref{fig:fig4}a we compare the lower Hubbard bands with the hole quasi-particle peak in the photo-doped state with hole doping level $n_h$ to the corresponding equilibrium spectrum in a purely hole-doped system with filling $n=1-n_h$. Interestingly, if the temperature $T$ of the equilibrium system is chosen equal to the temperature of the photo-doped system ($1/T=\tilde \beta=13$), the quasiparticle peak is overestimated. This can be explained by the fact that the presence of electron-like quasi-particles in the photo-doped state provides an additional scattering channel which drives the holes farther from the Fermi liquid regime into a resilient quasiparticle regime. Heuristically, this is confirmed by the observation that the quasiparticle peak of the transient state can be more accurately reproduced when a higher temperature, $\beta=7.1$, is chosen for the equilibrium system. In the strong-coupling regime considered in this letter, this additional scattering channel can be partially attributed to a \emph{repulsive} super-exchange interaction for neighboring electron and hole-like quasiparticles \cite{Li2019,Note1}. The analogous behavior is seen for the electron doped side (Fig.~\ref{fig:fig4}b), where we compare the upper Hubbard band region in the transient state and in an electron-doped system with filling $n=1+n_e$.

Another useful comparison between chemically doped and photo-doped states can be made for the optical conductivity $\sigma(\omega)$, Fig.~\ref{fig:fig4}c. The optical conductivity is evaluated as in Ref.~\cite{Li2019} and for more details see appendix. We mainly focus on the low frequency Drude peak. The mutual influence of electron and hole carriers becomes evident because adding up the conductivities $\sigma_e$+$\sigma_h$ of an electron-doped system at $n=1+n_e$ and a hole-doped system at $n=1-n_h$ ($\beta=13$) overestimates the Drude contribution. This observation is in line with the fact that the presence of holes (electrons) reduces the conductivity of electrons (holes) by providing additional scattering.  Instead, the Drude peak is well reproduced by the Drude peak of a chemically doped system with filling $n=1+n_e+n_h$ and inverse temperature $\tilde \beta$, similar to the half-filled case \cite{Werner2019}.

Finally, we comment on the relaxation dynamics. The effective temperature $\tilde T=1/\tilde \beta$ of the transient state is plotted as a function of time in Fig.~\ref{fig:fig3}(b) for different values of $\delta$. After the quench, $\tilde T$ decreases rapidly, as energy is transferred to the bath. At $t\gtrsim 300$, after the appearance of the quasiparticle peak, the dynamics slows down, and  $\tilde T$ relaxes with a roughly  constant rate towards the bath temperature $T_f$. This slowdown of the relaxation, which becomes more pronounced in the underdoped regime (Fig.~\ref{fig:fig3}c) is a rather striking result. Possible explanations based on phase space arguments and the bath density of states (such as a phonon bottleneck \cite{Rameau2016}) are unlikely, because a simulation of the same dynamics in the weak-$U$ metallic regime, with the same bath coupling $\lambda$ and bath density of states, shows a rapid cooling of the electrons at  fast rate until the temperature $T_f$ is reached up to the numerical accuracy (see the curve labeled as $U=2$, which has been obtained with non-equilibrium DMFT and a second-order self-energy). The inefficient transfer of energy from the correlated electron liquid to the phonons is reminiscent of what has been observed in heavy fermion systems \cite{Demsar2003}, although in the present case the system is not yet a Fermi-liquid, and an understanding in terms of a kinetic picture is therefore naturally difficult. At present we have no analytical understanding for this numerical prediction, but it should be observable by monitoring the evolution of the distribution function in photoemission spectroscopy, and it has the rather profound consequence that good metallic Fermi liquid states are hard to reach by photo-doping even in clean systems without trapping of charge carriers by impurities or in polaronic and excitonic states.

{\textit{Conclusion--}}  In summary, we studied the long time dynamics of photo-doped Mott insulators coupled to a thermal reservoir. By photo-doping an initially doped system, one can generate a strange metal with unequal densities of electron-like and hole-like carriers. 
The following aspects may be of interest to future studies and experiments: (i) We find that the photo-doped conductivity is in agreement with the conductivity of the chemically doped system, as observed in Ref.~\cite{Petersen2017}.  (ii) Holes and electrons do have different spectral properties. It will be interesting to see whether this behavior has an influence on the heat and charge transport. An intriguing experiment which would be within reach would be a diffusion measurement in a cold atom setting, similar to Ref.~\cite{Brown2019,Xu2019}. Excited states with doublons and holes can easily be generated by lattice shaking.  (iii) Finally, one remarkable finding is the extremely slow energy relaxation in the resilient quasiparticle regime. On the timescales of the simulation ($2000$ hopping times $\hbar/J$, corresponding to $5$ picoseconds if the bandwidth is $4J=1$eV) the effective temperature of the system remains in the range where the equilibrium system would be in the resilient quasiparticle regime. As a consequence of the slow relaxation, even on the ps timescale it is difficult to turn a 
photo-doped Mott insulator into a good metal. Alternative routes may be cooling by-photo-doping protocols \cite{Werner2019}. Our theoretical prediction of slow energy relaxation can be measured experimentally using cuprates and organic charge-transfer salts. 

This work was supported by the ERC Starting Grant No. 716648. The calculations 
have been done at the RRZE of the University Erlangen-Nuremberg. 
PW acknowledges support from ERC Consolidator Grant No. No. 724103.

\section{Appendix}

\subsection{Memory truncation of Kadanoff-Baym equations:}

The real-time evolution of correlated metallic states, starting from the initial
excitation to the formation of quasi-particle states is challenging with the
available numerical techniques \cite{Aoki2014}. Since the unitary time
evolution involves the non-Markovian time
propagation scheme, it limits in practice the maximum simulation
time. Recently, the truncation of memory times within non-equilibrium dynamical
mean-field theory has been explored \cite{Schueler2018}, which provides a strategy
for the simulation of long-time dynamics of photo-excited states.
In this scheme, the time evolution of quantum systems is computed on the L-shaped Kadanoff-Baym contour,
but the Kernel of the Kadanoff-Baym equations is truncated at some maximum relative time $t_c$.
This truncation significantly reduces
the computational cost required for long time dynamics, and enables to access
the thermalization of photo-excited states.

The local interacting lattice Green's function within the non-equilibrium dynamical mean-field theory
is calculated by solving the effective impurity problem with action
\begin{equation*}
S_\text{eff}=\int_C dt \left[\mu \sum_{\sigma}n_{\sigma}(t) + 
U n_{\uparrow}(t) n_{\downarrow}(t)\right] + \int_C dt dt' \Delta(t,t')
\end{equation*}
on the L-shaped Kadanoff-Baym contour. Here, we use the strong coupling
perturbative method (non-crossing approximation) as an impurity solver. Since the hybridization function
is an input to the impurity solver, the memory truncation in the Kadanoff-Baym equations has been introduced at
the level of $\Delta(t,t')$ \cite{Schueler2018}.

\begin{figure}[tbp]
\includegraphics[scale=1.2]{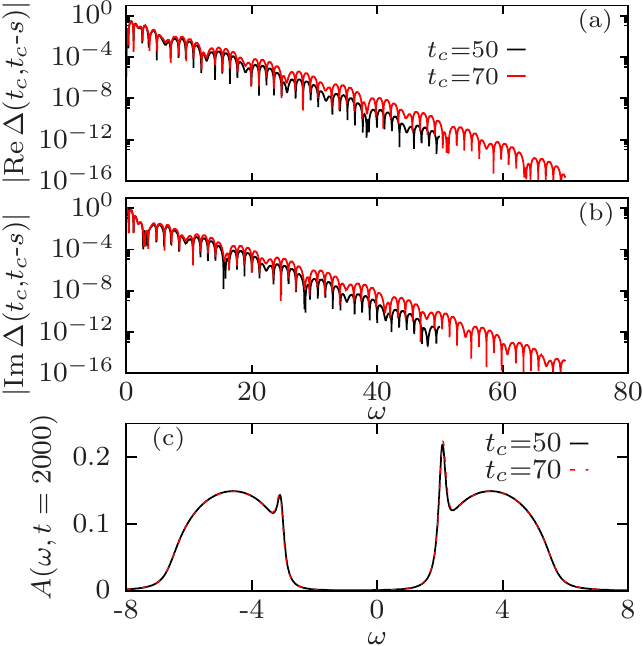}
\caption{The real part of $\Delta$ (a) and imaginary part of $\Delta$ (b) are
plotted for two different cut-off times and $\delta$=0.03. The corresponding
spectral functions 
are plotted in panel (c).
}
\label{fig:fig5}
\end{figure}

The results presented in the main paper are obtained with a
truncation of $\Delta(t,t')$ at $t_c=50$. A priori we do not know whether a certain
truncation time is sufficient and convergence with respect to  $t_c$ needs to be checked.
For a sufficiently large cutoff, the long time dynamics should not change upon a further increase in $t_c$.
The convergence test for $t_c=50$ is illustrated in Fig.~\ref{fig:fig5}.
The real and imaginary parts of the hybridization functions, plotted
in Fig.~\ref{fig:fig5}a) and b), show an exponential
decay for both cut-off times. 
In Fig.~\ref{fig:fig5}c) we plot the corresponding spectral functions, 
which show a good agreement. These results confirm that for the purpose of the
present analysis $t_c=50$ is sufficient.

\begin{figure}[tbp]
\includegraphics[scale=1.2]{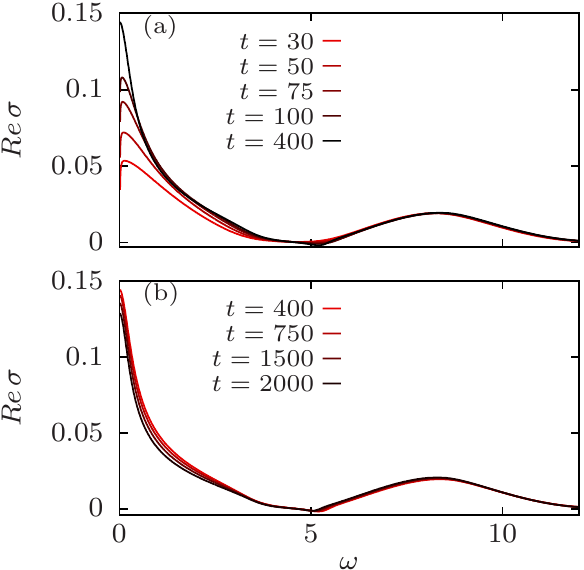}
\caption{The optical conductivity of the transient state is
plotted at early times in (a) and for long times in (b).}
\label{fig:fig6}
\end{figure}

\subsection{Optical conductivity:}
We have evaluated the optical conductivity of the transient state by
applying a Gaussian electric field pulse $E_{t'}(t)$ centered at
time $t'=t$ which couples to the hopping integral 
via a Peierls phase. 
The electric pulse adds a complex phase factor to the hybridization
function $\Delta(t,t')$=$\frac{J^2}{2}\left[e^{i\phi(t)}G(t,t')
e^{-i\phi(t)}+e^{-i\phi(t)}G(t,t')e^{i\phi(t)}\right]$ with a phase
$\phi(t)=-\int^{t}_0 ds E(s)$ 
\cite{Werner2019,Li2019}. The external field
induces a finite current in the transient
state and is measured as $I(t)$=$\text{Re}\left[G*\Delta_{L-R}\right]^{<}(t,t)$,
with $\Delta_{L-R}(t,t')$=$\frac{J^2}{2}\left[e^{i\phi(t)}G(t,t')
e^{-i\phi(t)}-e^{-i\phi(t)}G(t,t')e^{i\phi(t)}\right]$. The frequency-dependent
optical conductivity is given by $\text{Re}\hspace{0.05cm}\sigma(\omega,t)=I_{t'}(\omega)/E_{t'}(\omega)$
where $I_{t'}(\omega)$, $E_{t'}(\omega)$ are the Fourier transform of the induced
current and electric
pulse, respectively.

The optical conductivity of the transient state is plotted
in Fig.~\ref{fig:fig6} at different times for $\delta=0.03$. As observed
in the spectral function, the coherent peak in the optical conductivity,
called Drude peak, is emerging after the temperature quench (see Fig.~\ref{fig:fig6}a),
and it is fully established on the time scale on which quasi-particles in the
spectral functions are completely formed. After that, the Drude peak
evolves slowly (see Fig.~\ref{fig:fig6}b) due to the recombination of
photo-excited charge carriers. Apart from the coherent peak, the optical conductivity
has an incoherent peak at $\omega \sim 8$, which is corresponding to charge
excitations between the lower and upper Hubbard band.

\bibliographystyle{apsrev4-1}
\bibliography{apssamp}

\jl{\footnotetext[1]{The interaction has the form of $-J_{\rm ex}(n_i-1/2)(n_j-1/2)$ with $J_{\rm ex}=4J^2/U$. Starting with neighboring doubly occupied (electron-like) and empty (hole-like) lattice sites, the interaction emerges due to a virtual process forming a virtual intermediate state of two singly occupied sites. The interaction is repulsive since the intermediate state has lower energy ($-U$) than the initial and final real states. }}

\end{document}